\documentclass[%
 reprint,
 amsmath,amssymb,
 aps,
]{revtex4-1}

\usepackage{graphicx}
\usepackage{dcolumn}
\usepackage{bm}

\begin{document}

\preprint{APS/123-QED}

\title{Logic gates on stationary dissipative solitons}

\author{Bogdan~Kochetov$^1$}
\author{Iaroslavna~Vasylieva$^2$}
\author{Alexander~Butrym$^3$}
\author{Vladimir~R.~Tuz$^{1,2}$}
\email{tvr@jlu.edu.cn; tvr@rian.kharkov.ua}
\affiliation{$^1$State Key Laboratory of Integrated Optoelectronics, College of Electronic Science and Engineering, \\ International Center of Future Science, Jilin University, 2699 Qianjin St., Changchun 130012, China}
\affiliation{$^2$Institute of Radio Astronomy of National Academy of Sciences of Ukraine, 4, Mystetstv St., Kharkiv 61002, Ukraine} 
\affiliation{$^3$Department of Theoretical Radio Physics, V.~N.~Karazin Kharkiv National University, 4, Svobody Sq., Kharkiv 61022, Ukraine}

\begin{abstract}
Stable dissipative solitons are perfect carries of optical information due to remarkable stability of their waveforms that allows the signal transmission with extremely dense soliton packing without loosing  the encoded information. Apart of unaffected passing of solitons through a communication network, controllable transformations of soliton waveforms are needed to perform all-optical information processing. In this paper we employ the basic model of dissipative optical solitons in the form of the complex Ginzburg-Landau equation with a potential term to study the interactions between two stationary dissipative solitons being under the control influences and use those interactions to implement various logic gates. Particularly, we demonstrate NOT, AND, NAND, OR, NOR, XOR, and XNOR gates, where the plain (fundamental soliton) and composite pulses are used to represent the low and high logic levels. 
\end{abstract}

\maketitle

\section{\label{Intro}Introduction}
Dissipative optical solitons, localized waves in non-integrable systems far from equilibrium whose properties depend dramatically on the internal energy balance, were realized at the beginning of 1990s \cite{Picholle_PRL_1991,Vanin_PRA_1994}. Due to the subsequent theoretical and experimental studies of these solitary waves and unification of their features the ideas of self-organization, common for the animate and inanimate worlds, were elaborated in the development of concept of dissipative solitons \cite{Akhmediev_Book1,Akhmediev_Book2,Liehr_Book}. The theoretical framework for study of dissipative solitons is based on the complex Ginzburg-Landau equation (CGLE), which accounts for the supply and absorption of energy in presence of nonlinear and dispersive (diffractive) environment, crucially important conditions for the development of localized dissipative structures \cite{Cross_RMP_1993,Aranson_RMP_2002,García-Morales_CP_2012}. The CGLE admits a few classes of stable solutions representing the rich variety of dissipative solitons and nontrivial behaviour of their evolution  \cite{Akhmediev_Chapter_2005}. In fact, the localized waves governed by the one-dimensional CGLE can evolve as solitons with stationary \cite{Malomed_PD_1987, Malomed_PRA_1990, Fauve_PRL_1990, van_Saarloos_PD_1992, Afanasjev_PRE_1996, Renninger_PRA_2008}, periodically, quasi-periodically, and aperiodically (chaotically) pulsating waveforms \cite{Deissler_PRL_1994, Soto-Crespo_PRL_2000, Akhmediev_PRE_2001}, moving pulses\cite{Afanasjev_PRE_1996}, exploding solitons \cite{Soto-Crespo_PRL_2000, Akhmediev_PRE_2001, Soto-Crespo_PLA_2001, Cundiff_PRL_2002, Descalzi_PRE_2011}, solitons with periodical and chaotic spikes of extreme amplitude and short duration \cite{Chang_OL_2015, Chang_JOSAB_2015, Soto-Crespo_JOSAB_2017}, multisoliton solutions \cite{Akhmediev_PRL_1997}, and in the form of stable dynamic bound states \cite{Turaev_PRE_2007}. Remarkably, these different forms of dissipative solitons coexist to each other when the equation coefficients belong to certain regions \cite{Afanasjev_PRE_1996,Soto-Crespo_PRL_2000,Akhmediev_PRE_2001,Soto-Crespo_PLA_2001,Descalzi_PhysicaA_2006}. Moreover, the basic CGLE can easily be extended to more general models accounting for the impact of such high-order effects as third-order dispersion, fourth-order spectral filtering, self-stepping, and stimulated Raman scattering \cite{Soto-Crespo_PRE_2002,Achilleos_PRE_2016,Sakaguchi_OL_2018,Uzunov_PRE_2018,Gurevich_2019} as well as an external control \cite{Boardman_Chapter_2005,Boardman_2006}. In fact, more specific models have been used to study the turbulent-like intensity and polarization rogue waves in a Raman fiber laser \cite{Sugavanam_LPR_2015}, stationary solitary pulses in a dual-core fiber laser \cite{Malomed_C_2007}, the interaction of stationary, oscillatory and exploding counter-propagating dissipative solitons \cite{Descalzi_EPJ_2015,Descalzi_C_2018}, the existence of stable three-dimensional dissipative localized structures in the output of a laser coupled to a distant saturable absorber \cite{Javaloyes_PRL_2016}, the emergence and the stability of temporally localized structures in the output of a semiconductor laser passively mode locked by a saturable absorber in the long-cavity regime \cite{Schelte_PRA_2018}, and dissipative solitons in Bose-Einstein condensates \cite{Malomed_Book,Wouters_PRL_2007,Ostrovskaya_PRA_2012,Xue_PRL_2014,Smirnov_PRB_2014,Xue_OE_2018}.

The significant stability of dissipative solitons with respect to the distortion effects allows the soliton passing with very dense pulse packing without loosing the encoded information that makes them ideal carriers of information in new optical systems. The development of such systems for performing the all-optical information processing requires robust devices on the dissipative optical solitons, similarly to those on the conservative ones \cite{Stegeman_Science_1999}. First of all the devices can be implemented in the framework of the soliton-soliton interactions. Particularly, the AND and OR logic gates based on the self-interactions of bright dissipative polariton solitons have theoretically been demonstrated in \cite{Cancellieri_PRB_2015}. On the other hand, the interactions of dissipative solitons can explicitly be controlled by the externally applied influence. In fact, this control has repeatedly been added to equations governing the soliton dynamics in the form of an external potential. Particularly, the diffusion-induced turbulence has been modelled on the base of the CGLE with an additional term accounting for the global delayed feedback \cite{Battogtokh_PD_1996} and a gradient force \cite{Xiao_PRL_1998}. Spatial localization and dynamical stability of Bose-Einstein condensates of exciton-polaritons in microcavities are examined in \cite{Ostrovskaya_PRA_2012,Xue_PRL_2014}. The nonlinear Schr\"odinger equation with a longitudinal defect \cite{Fratalocchi_PRE_2006}, an external delta potential \cite{Holmer_JNS_2007}, and a longitudinal potential barrier \cite{Yang_OE_2008} appear in optical applications for beam splitters. The complex dynamics of dissipative solitons in active bulk media with spatially modulated refractive indexes in the form of a sharp potential barrier \cite{He_JOSAB_2010}, umbrella-shaped \cite{Yin_JOSAB_2011}, and radial-azimuthal \cite{Liu_OE_2013} potentials has been studied on the base of the one- and two-dimensional cubic-quintic CGLEs.

Accounting external magnetic field in nonlinear magneto-optic waveguides leads to another example of controllable optical solitons \cite{Boardman_1997}. Due to the applied magnetic field the time reversal symmetry is locally broken that leads to significantly different propagation conditions of counter-propagating dissipative optical solitons whose envelops are governed by the cubic-quintic CGLE with a potential term \cite{Boardman_Chapter_2005, Boardman_2006}. Recently this robust model has successfully been used to perform a selective lateral shift within a group of stable noninteracting fundamental dissipative solitons \cite{OptLett_2017}, to replicate dissipative solitons and vortices \cite{PRE_2017,PRE_2018}, and to induce the waveform transitions between different dissipative solitons \cite{Chaos_2018,PD_2019}.

Since one of the central challenges in the development of promising optical systems based on stable dissipative solitons is the getting of full control over soliton interactions, we further employ the one-dimensional cubic-quintic CGLE with a potential term to implement logic gates on two different stationary dissipative solitons. Each of these logic gates operates due to a specific control potential applied locally along the propagation distance.  

The rest of the paper is organized as follows. In Section~\ref{MathMod} we introduce the basic mathematical model of dissipative solitons in the form of one-dimensional cubic-quintic CGLE with a potential term. This model supports coexistence of two stationary dissipative solitons (plain and composite pulses) with significantly different waveforms and spectra as well as describes their interactions under the control of applied potential. In Section~\ref{LG}, having applied appropriate control potentials we demonstrate NOT, AND, NAND, OR, NOR, XOR, and XNOR gates on dissipative solitons, where the plain and composite pulses represent the low and high logic levels. Conclusions and remarks finalize the paper in Section~\ref{Concl}.

\section{\label{MathMod}Mathematical Model of Controllable Dissipative Solitons}
The cubic-quintic CGLE supplemented by a potential term with an explicit coordinate dependence forms the background for simulations of dissipative solitons in many optical applications. Particularly, this equation appears in the theory of planar nonlinear magneto-optic waveguides \cite{Boardman_Chapter_2005,Boardman_2006,OptLett_2017} and describes the evolution of electromagnetic fields in nonlinear optical media with spatially modulated refractive index \cite{Yang_OE_2008,He_JOSAB_2010,Yin_JOSAB_2011,Liu_OE_2013}. Here, we adopt the notations used in optics and write down the CGLE in the following form
\begin{align}
\label{CQCGLE}
\mathrm{i}&\frac{\partial\Psi}{\partial z}+\mathrm{i}\delta\Psi+\left(\frac{1}{2}-\mathrm{i}\beta\right)\frac{\partial^2\Psi}{\partial x^2} \nonumber+\left(1-\mathrm{i}\varepsilon\right)\left|\Psi\right|^2\Psi\\ &-\left(\nu-\mathrm{i}\mu\right)\left|\Psi\right|^4\Psi + Q(x,z)\Psi = 0,
\end{align}
where $\Psi\left(x,z\right)$ is the complex slowly varying envelop of the transverse $x$ and longitudinal $z$ coordinates. All coefficients of Eq.~\eqref{CQCGLE} are assumed to be positive quantities. It implies that $\delta$ and $\beta$ account for the linear absorption and diffusion, $\nu$ stands for the self-defocusing effect due to the quintic nonlinearity, while $\varepsilon$ and $\mu$ are the cubic gain and quintic loss coefficients, respectively.

The potential $Q(x,z)$ accounts for the influence of linear conservative forces applied externally to control the evolution of complex envelop $\Psi(x,z)$. Its particular spatial distribution depends on the physical origin of applied forces. For example, in some optical applications the potential $Q(x,z)$ can account for the linear magneto-optic effect \cite{Boardman_Chapter_2005,Boardman_2006} and spatial modulation of the refractive index \cite{He_JOSAB_2010, Yin_JOSAB_2011,Liu_OE_2013,Liu_OL_2010, Liu_OE_2011}. For control purposes it is logical to assume that the potential acts locally along the propagation distance having a finite supporter along the $z$ axis. Without loss of generality we choose the longitudinal dependence of potential in the form of a piecewise constant function and write down the potential as follows
\begin{equation}
\label{Q}
Q(x,z) = \sum_{i=1}^{N} q_i(x)\left[h(z-a_i)-h(z-b_i)\right],
\end{equation}
where $N$ is the number of control manipulations, $q_i(x)$ is the transverse variation of the potential during the $i$-th control manipulation, $h(\cdot)$ is the Heaviside step function, and $a_i<b_i$ are some points on the $z$ axis at which the potential changes its transverse distribution. The transverse profiles $q_i(x)$, end points $a_i$, $b_i$, and number $N$ should be chosen to perform certain control over soliton waveforms. Here, they are specified to implement logic gates as discussed later on.

The CGLE admits existence of a few different attractors at the same values of its coefficients that means coexistence of different stable dissipative solutions for a given set of parameters \cite{Afanasjev_PRE_1996,Soto-Crespo_PRL_2000,Akhmediev_PRE_2001,Soto-Crespo_PLA_2001,Descalzi_PhysicaA_2006}. Particularly, in wide regions of the parameter space the CGLE allows coexistence of two stable stationary dissipative solitons with different waveforms \cite{Afanasjev_PRE_1996}. One of them is the fundamental soliton, which is also called the \textit{plain} pulse, while another one is the so-called \textit{composite} pulse. Fig.~\ref{fig_1} shows the typical intensity distributions and normalized power spectra of these two solitons, where the dash-dot red and solid blue lines indicate the plain and composite pulses, respectively. Moreover, Fig.~\ref{fig_1} shows the energies of both solitons calculated in the coordinate space ($E_{pp}$, $E_{cp}$) and in the Fourier domain ($\hat{E}_{pp}$, $\hat{E}_{cp}$), where the subindexes stand for the pulse abbreviations. To plot the waveforms of coexisting solitons [Fig.~\ref{fig_1}(a)] and their spectra [Fig.~\ref{fig_1}(b)] at some point $z=z_0$ we numerically solve Eq.~\eqref{CQCGLE} using the following set of coefficients $\delta = 0.5$, $\beta = 0.5$, $\mu = 1$, $\nu = 0.1$, and $\varepsilon = 2.52$ and having applied zero potential, i.e. $Q(x,z)=0$. The same set of numerical values for the coefficients of Eq.~\eqref{CQCGLE} we use here in all our numerical simulations, which are performed using the exponential time differencing method as well as its Runge-Kutta modification of second- and fourth-order accuracy in the Fourier domain \cite{Cox_2002}. We apply the fast Fourier transform to the complex amplitude $\Psi(x,z)$ with respect to the transverse coordinate $x$ transforming it to its Fourier amplitude $\hat{\Psi}(k_x,z)$. This imposes the periodic boundary condition
\begin{equation}
\label{BC}
\Psi(x,z) = \Psi(x+L_x,z),~~~~\forall(x,z)\in\mathbb{R}\times[0,+\infty),
\end{equation}
with some period $L_x>0$. Therefore, the computational domain is reduced to the finite rectangular $[-L_x/2,L_x/2]\times[0,L_z]$, where its width $L_x=100$ is chosen to ensure that all non-negligible parts of waveforms are within the domain, while its length $L_z$ is chosen to ensure the completion of simulations. Typically, it varies in the range $900\leq L_z\leq3000$. We sample the computational domain with $N_x=2^{10}$ points along the transverse coordinate $x$ and use the step $\Delta z=10^{-3}$ to discretize  the domain along the longitudinal coordinate $z$. 

Being stationary solutions to Eq.~\eqref{CQCGLE}, the plain and composite pulses can easily be excited by numerous appropriate waveforms used as initial conditions. In fact, for the specified coefficients of Eq.~\eqref{CQCGLE} the plain and composite pulses quickly develop from the initial waveforms $\Psi_{pp}(x)$ and $\Psi_{cp}(x)$, which are respectively defined as
\begin{equation}
\label{IC}
\Psi_{pp}(x)=\mathrm{sech}(x),~~~~\Psi_{cp}(x)=\exp\left(-\frac{x^2}{25}\right).
\end{equation}

\section{\label{LG}Logic Gates}
\begin{figure}[htbp]
\centering
\includegraphics[width=\linewidth]{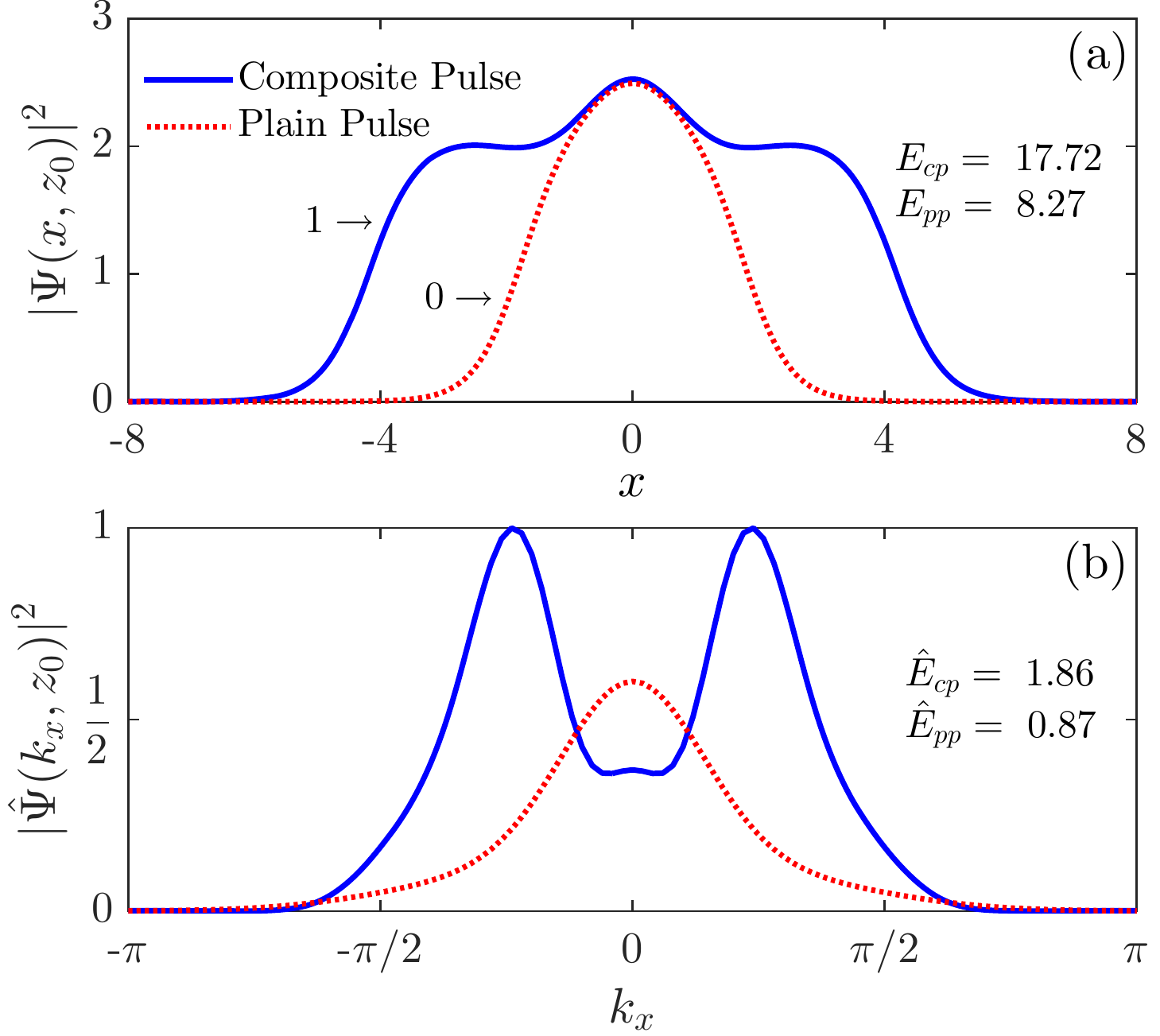}
\caption{Two stable stationary dissipative solitons coexisting for the same set of equation coefficients ($\delta = 0.5$, $\beta = 0.5$, $\mu = 1$, $\nu = 0.1$, and $\varepsilon = 2.52$) represent low $(0)$ and high $(1)$ logic levels: (a) intensities; (b) normalized power spectra.}
\label{fig_1}
\end{figure}

In this section we demonstrate the implementation of logic gates on dissipative solitons in our simulations based on the numerical analysis of the model \eqref{CQCGLE}-\eqref{BC}. Particularly, we exploit two stable stationary solitons admitted by Eq.~\eqref{CQCGLE} to represent logic levels. To be specific, we assume that the plain pulse represents the low ($0$) logic level, while the composite pulse represents the high ($1$) level. These pulses have different waveforms [Fig.~\ref{fig_1}(a)] and strictly distinguished spectra [Fig.~\ref{fig_1}(b)]. Moreover, the plain and composite pulses can also be considered as two isolated stable fixed points (attractors) in an infinite-dimensional phase space of the system~\eqref{CQCGLE} \cite{Akhmediev_Chapter_2005}. It means that the waveforms and spectra of the pulses are unchangeable along the propagation distance as long as the equation coefficients are fixed and the potential is not applied. In other words, there can be no uncontrollable overlap between the plain and composite pulses that allow them to represent logic levels ideally.

On the other hand, having applied the external potential~\eqref{Q} one can get control over the soliton waveforms, for example, to transit the plain pulse to the composite pulse and to return its waveform back \cite{PD_2019}. In general, such transitions induced by an external potential can be possible between an arbitrary pair of stable coexisting dissipative solitons if an appropriate control potential is applied \cite{Chaos_2018,PD_2019}. We can imagine a particular waveform transition as an forced displacement of a point in the phase space from a basin of attraction of given attractor to a vicinity of another attractor. Here we further elaborate the ideas of induced waveform transitions \cite{OptLett_2017,PRE_2017,Chaos_2018,PRE_2018,PD_2019} in the development of controllable interaction between the plain and composite pulses that finally lead us to the implementation of all the logic gates. Below we subsequently demonstrate the most important of them.

\subsection{\label{NOT}NOT gate}
\begin{figure*}[htbp]
\centering
\includegraphics[width=1\linewidth]{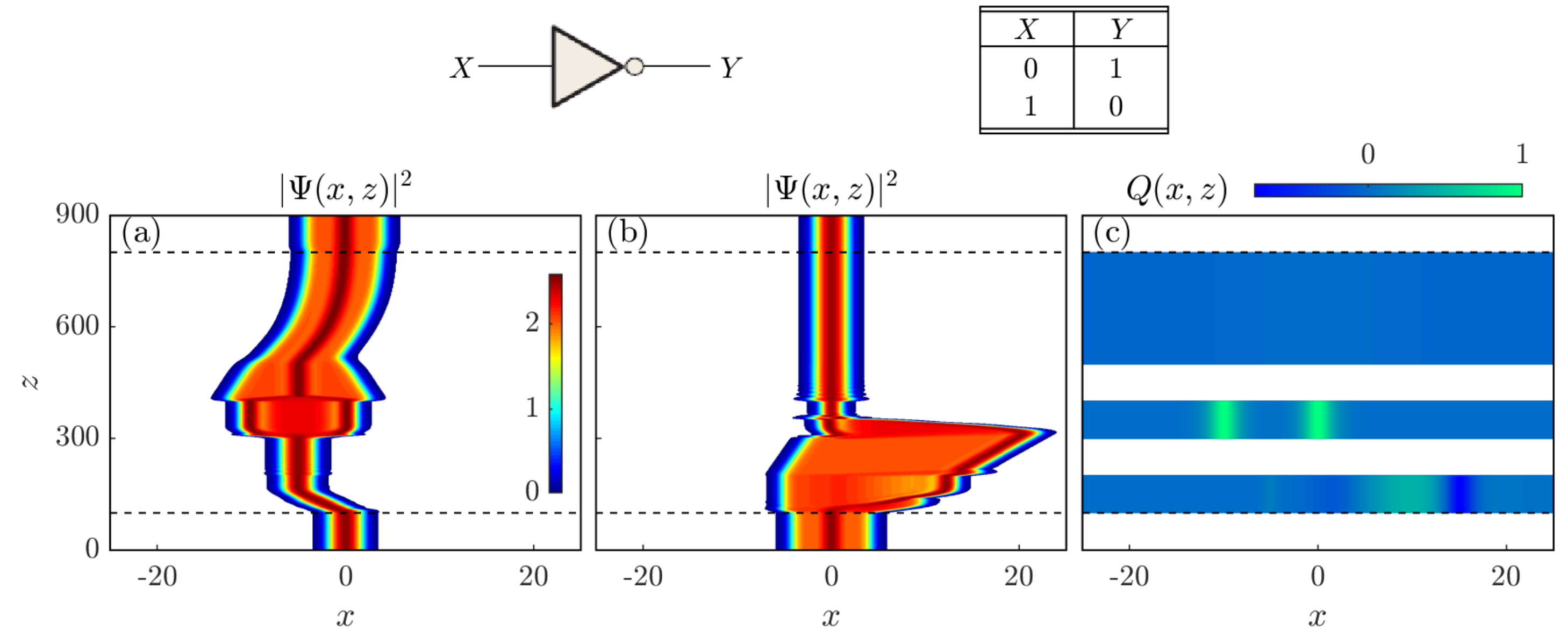}
\caption{NOT gate. Truth table for a NOT gate and its implementation on the dissipative soliton bits (a), (b) due to externally applied potential $Q(x,z)$ with parameters \eqref{q_not} (c).}
\label{fig_2}
\end{figure*}

We start with the consideration of a NOT gate, which performs the operation of inversion changing one logic level to the opposite level. In terms of the introduced soliton bits, it changes the plain pulse to the composite pulse [Fig.~\ref{fig_2}(a)] and the composite pulse to the plain pulse [Fig.~\ref{fig_2}(b)]. In fact, to complete the description of NOT gate in the framework of the model~\eqref{CQCGLE}-\eqref{BC} we have to specify the unknown number of manipulations $N$, start and end points $a_i$ and $b_i$, and the transverse dependencies $q_i(x)$ in the potential~\eqref{Q} used to implement the NOT gate. We should note that these parameters chosen in different way can lead to multiple implementations of the NOT gate, i.e. there is no unique choice for them. Particularly, we assume that each of the transverse dependencies $q_i(x)$ is a sum of a few scaled $\mathrm{sech}(x)$ functions, which are used as trial functions to approximate the transverse dependence of potential. For example, the potential~\eqref{Q} suitable to implement the NOT gate contains three control manipulations $N=3$ with the following parameters

\begin{align}
\label{q_not}
a_i\in\{100,300,500\},~~~b_i\in\{200,400,800\},\nonumber\\
q_1(x)=\frac{1}{10}\mathrm{sech}(2x+10) -\frac{1}{4}\mathrm{sech}\left(\frac{x-2}{2}\right)\nonumber\\ +\frac{1}{2}\mathrm{sech}\left(\frac{x-10}{4}\right) -\mathrm{sech}(x-15),\nonumber\\ q_2(x)=\mathrm{sech}(x+10)+\mathrm{sech}(x),\nonumber\\ q_3=\frac{1}{20}\left[\mathrm{sech}\left(\frac{x}{5}\right)\right.\nonumber\\ \left.-\mathrm{sech}\left(\frac{x+19}{10}\right) -\mathrm{sech}\left(\frac{x-19}{10}\right)\right].
\end{align}
Two-dimensional spatial distribution of the potential~\eqref{Q} with parameters~\eqref{q_not} is plotted in Fig.~\ref{fig_2}(c), while the evolution of the plain and composite pulses under its influence is shown in Fig.~\ref{fig_2}(a) and Fig.~\ref{fig_2}(b) respectively, where one can see two-dimensional intensity plots of complex envelops $|\Psi(x,z)|^2$ of these pulses. More precisely, in Fig.~\ref{fig_2}(a) we see that the input plain pulse is transited by potential~\eqref{Q}, \eqref{q_not} to the output composite pulse, while according to Fig.~\ref{fig_2}(b) we conclude that the same potential transits the input composite pulse to the output plain pulse. Thereby potential~\eqref{Q} with parameters~\eqref{q_not} produces the inverted output pulse with respect to the given input pulses. This operation is summarized in the table inserted in Fig.~\ref{fig_2}, where $X$ and $Y$ stand for the input and output soliton bits, respectively. This table is just the truth table for a NOT gate.

As mentioned above, potential~\eqref{Q}, \eqref{q_not} inverts the input soliton bits performing three subsequent control stages, which are clearly seen in Fig.~\ref{fig_2}(c). At the first stage, the applied potential has the asymmetrical transverse distribution $q_1(x)$, which is chosen to act selectively on the input plain and composite pulses whose waveforms are initially centered at $x=0$. In fact, it shifts the plain pulse along the negative direction of the $x$ axis [Fig.~\ref{fig_2}(a)] and changes the composite pulse stretching its waveform along the opposite direction [Fig.~\ref{fig_2}(b)]. Between the first and second stages, the shifted plain pulse keeps its position unchanged [Fig.~\ref{fig_2}(a)], while both fronts of the perturbed composite pulse move along the positive direction of the $x$ axis [Fig.~\ref{fig_2}(b)] that leads to stronger spatial separation of pulses. At the second stage, the two-peaked symmetric potential $q_2(x)$ transits the shifted plain pulse to the composite pulse [Fig.~\ref{fig_2}(a)], while the perturbed composite pulse is transited to the plain pulse [Fig.~\ref{fig_2}(b)]. Between the second and third stages, the pulses released of the potential influence evolve gradually to their unperturbed waveforms. Finally, at the third stage, the relatively weak potential with symmetrical profile $q_3(x)$ is applied to shift the peak of inverted plain pulse at the initial point $x=0$ [Fig.~\ref{fig_2}(a)]. The waveform of inverted composite pulse was centered around the point $x=0$ during the second stage. Therefore, its position is not changed during the third stage [Fig.~\ref{fig_2}(b)]. We should note that the third stage is the longest one because the lateral shifting of composite pulse [Fig.~\ref{fig_2}(a)] can only be performed by a weak attractive potential, which slowly shifts the composite pulse without considerable squeezing of its waveform. Otherwise, being attracted by a strong potential the composite pulse can collapse to the plain pulse \cite{PD_2019}.

\subsection{\label{AND}AND and NAND gates}
\begin{figure*}[htbp]
\centering
\includegraphics[width=\linewidth]{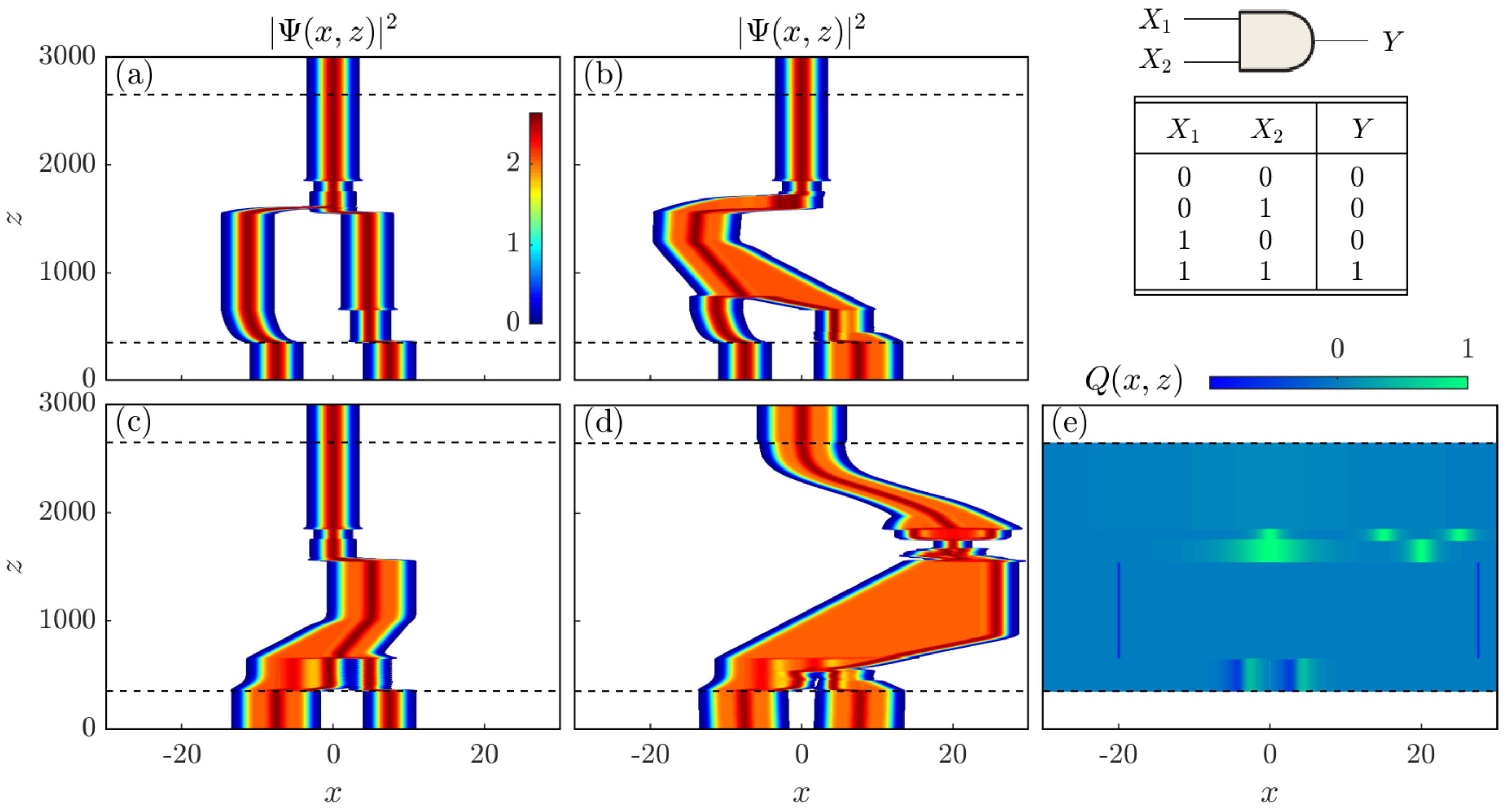}
\caption{AND gate. Truth table for a 2-input AND gate and its implementation on the dissipative soliton bits (a)-(d) due to externally applied potential $Q(x,z)$ with parameters \eqref{q_and} (e).}
\label{fig_3}
\end{figure*}

Now we consider the implementation of an AND gate with two inputs using the controllable model~\eqref{CQCGLE}-\eqref{BC}, which supports the dissipative soliton bits in a form of the plain and composite pulses (Fig.~\ref{fig_1}). The AND gate produce the composite pulse output only when all of the inputs are the composite pulses [Fig.~\ref{fig_3}(d)]. When any of the inputs is the plain pulse, the output is the plain pulse [Figs.~\ref{fig_3}(a)-(c)]. In fact, the implementation of the 2-input AND gate presented in Fig.~\ref{fig_3} is performed using potential~\eqref{Q} with the following parameters
\begin{align}
\label{q_and}
a_i\in\{350,650,1550,1750,1850\},\nonumber\\
b_i\in\{650,1550,1750,1850,2650\},\nonumber\\
q_1(x)=\frac{\mathrm{tanh}(x+3.5)}{\mathrm{cosh}(x+3.5)} + \frac{\mathrm{tanh}(x-3.5)}{\mathrm{cosh}(x-3.5)},\nonumber\\
q_2(x)=-\mathrm{sech}\left(10x+200\right)-\mathrm{sech}\left(10x-275\right),\nonumber\\
q_3(x)=\mathrm{sech}\left(\frac{x}{3}\right)+\mathrm{sech}(x-20),\nonumber\\
q_4(x)=\mathrm{sech}(x)+\mathrm{sech}(x-15)+\mathrm{sech}(x-25),\nonumber\\
q_5(x)=\frac{1}{20}\mathrm{sech}\left(\frac{x}{10}\right) +\frac{1}{25}\mathrm{sech}\left(\frac{x}{5}\right).
\end{align}

\begin{figure*}[htbp]
\centering
\includegraphics[width=\linewidth]{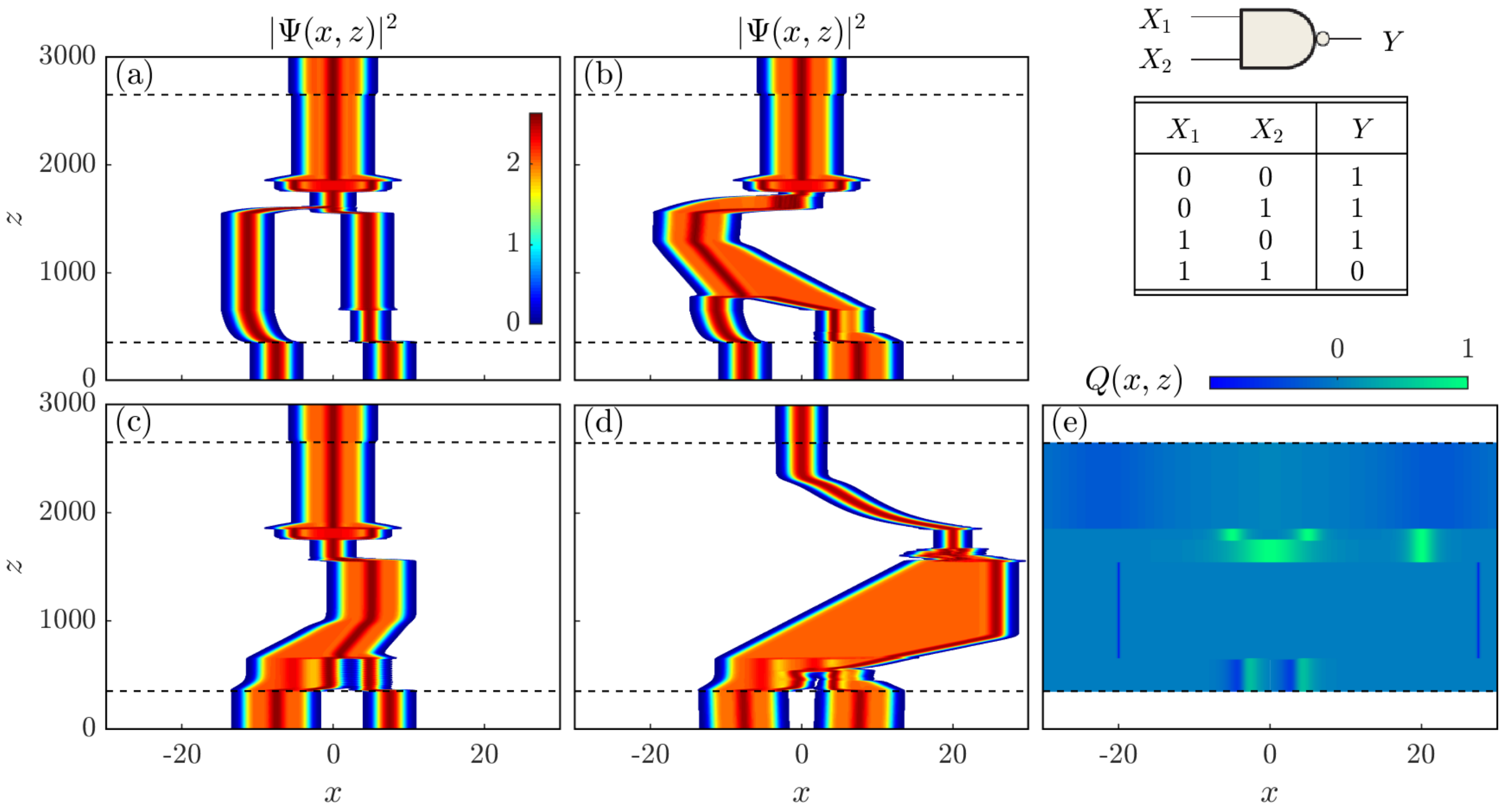}
\caption{NAND gate. Truth table for a 2-input NAND gate and its implementation on the dissipative soliton bits (a)-(d) due to externally applied potential $Q(x,z)$ with parameters \eqref{q_and}, where the last two control manipulations are replaced by Eqs.~\eqref{q_nand} (e).}
\label{fig_4}
\end{figure*}

Figs.~\ref{fig_3}(a)-(d) show the two-dimensional intensity plots $|\Psi(x,z)|^2$ representing the waveform evolution for all possible combinations of two input pulses, where the peaks of left and right input pulses are respectively located at the points $x=-7.5$ and $x=7.5$, while the output pulse is centered around the point $x=0$. In each of four cases presented in Figs.~\ref{fig_3}(a)-(d), the waveform evolution of two input pulses is controlled by potential~\eqref{Q} with the same parameters~\eqref{q_and} whose two-dimensional spatial distribution $Q(x,z)$ is plotted in Fig.~\ref{fig_3}(e). This control potential selectively transits the pairs of input pulses to certain single pulse in five stages ($N=5$) leading to the different output pulses depending on a particular combination of input pulses. Particularly, Fig.~\ref{fig_3}(a) shows the controllable evolution of two plain pulses to one plain pulse that corresponds to the first row in the truth table for a 2-input AND gate presented in Fig.~\ref{fig_3}. Moreover, in Figs.~\ref{fig_3}(b) and (c) we see how two different pairs comprised of the plain and composite pulses are transited to the plain pulse. These two transitions correspond to the second and third rows in the truth table in Fig.~\ref{fig_3}. Finally, in Fig.~\ref{fig_3}(d) the last possible input combination of two composite pulses is transited to the composite pulse corresponding to the fourth row in the truth table in Fig.~\ref{fig_3}.

In other words, in Figs.~\ref{fig_3}(a)-(d) we demonstrate the four basic rules for multiplying the soliton bits, where each multiplication is represented by the controllable interaction of two input pulses induced by the external potential~\eqref{Q}, \eqref{q_and}. Looking at Eqs.~\eqref{q_and} and Fig.~\ref{fig_3}(e) we see that this potential is chosen to perform the rules for five control manipulations over soliton waveforms. First of all we apply $q_1(x)$ from Eq.~\eqref{q_and} to perturb the waveforms of input pulses selectively, i.e. accounting for the soliton bit combination of input pulses. Then we apply the second manipulation $q_2(x)$ to release the selectively perturbed waveforms as well as to prevent the wide spreading of released waveforms along the transverse direction. As a result, the perturbed waveform of two input composite pulses is significantly moved along positive direction of the $x$ axis [Fig.~\ref{fig_3}(d)], while for other input pulses the waveforms are slightly shifted along positive direction of the $x$ axis [Fig.~\ref{fig_3}(c)] or are shifted in the opposite direction [Figs.~\ref{fig_3}(a), (b)] leading to the spatial separation of waveforms along the $x$ axis. Further we apply $q_3(x)$ to transit all the transformed and shifted waveforms to the plain pulses. However, the waveform evolved from two composite pulses is transited to the plain pulse whose peak is located at the point $x=20$ [Fig.~\ref{fig_3}(d)], while other waveforms are transited to the plain pulse with peak at $x=0$ [Figs.~\ref{fig_3}(a)-(c)]. The fourth control manipulation $q_4(x)$ is applied to transit the plain pulse centered around the point $x=20$ to the composite pulse [Fig.~\ref{fig_3}(d)] and to prevent the lateral shift of the plain pulses centered at the point $x=0$ [Figs.~\ref{fig_3}(a)-(c)]. The last manipulation $q_5(x)$ is applied to shift the composite pulse at the point $x=0$ [Fig.~\ref{fig_3}(d)], while the transverse positions of the plain pulses are not changed because they have already centered at the point $x=0$ [Figs.~\ref{fig_3}(a)-(c)]. Thus, we complete the implementation of the 2-input AND gate on the dissipative soliton bits.

Having replaced the last two control manipulations $q_4(x)$ and $q_5(x)$ in Eqs.~\eqref{q_and} by the following ones 
\begin{align}
\label{q_nand}
q_4(x)=\mathrm{sech}(x+5)+\mathrm{sech}(x-5)+\mathrm{sech}(x-20),\nonumber\\
q_5(x)=\frac{1}{25}\left[\mathrm{sech}(x) +\mathrm{sech}\left(\frac{x}{5}\right)\right] \nonumber\\ -\frac{3}{10}\left[\mathrm{sech}\left(\frac{x+22}{5}\right) +\mathrm{sech}\left(\frac{x-22}{5}\right)\right],
\end{align}
we get the appropriate potential to implement a 2-input NAND gate whose operation is opposite to that of the AND in terms of the output level. As it is summarized in the truth table presented in Fig.~\ref{fig_4}, for a 2-input NAND gate, output $Y$ is low only when inputs $X_1$ and $X_2$ are high; $Y$ is high when either $X_1$ or $X_2$ is low, or when both $X_1$ and $X_2$ are low. The implementation of the 2-input NAND gate on the dissipative soliton bits  we demonstrate in Figs.~\ref{fig_4}(a)-(d), where each graph shows the evolution of soliton intensity $|\Psi(x,z)|^2$ under the influence of control potential \eqref{Q} with the same parameters \eqref{q_and} used to implement the AND gate except for the last two manipulations $q_4(x)$ and $q_5(x)$, which we now take in the form of Eqs.~\eqref{q_nand}. Two-dimensional spatial distribution of this potential $Q(x,z)$ is plotted in Fig.~\ref{fig_4}(e).

We see that the graphs presented in Figs.~\ref{fig_4}(a)-(d) are similar to those shown in Figs.~\ref{fig_3}(a)-(d), respectively. Moreover, the first three control manipulations are identical for both gates. However, we apply the fourth control manipulation $q_4(x)$ to transit the plain pulses centered around the point $x=0$ to the composite pulse [Figs.~\ref{fig_4}(a)-(c)] and to prevent the waveform change of plain pulse centered at the point $x=20$ [Fig.~\ref{fig_4}(d)]. The last manipulation $q_5(x)$ is applied to shift the plain pulse laterally from the point $x=20$ to the point $x=0$ [Fig.~\ref{fig_4}(d)] and save the waveforms of composite pulses already centered at the point $x=0$ [Figs.~\ref{fig_4}(a)-(c)]. That implements the 2-input NAND gate based on the dissipative solitons.

\subsection{\label{OR}OR and NOR gates}
\begin{figure*}[htbp]
\centering
\includegraphics[width=\linewidth]{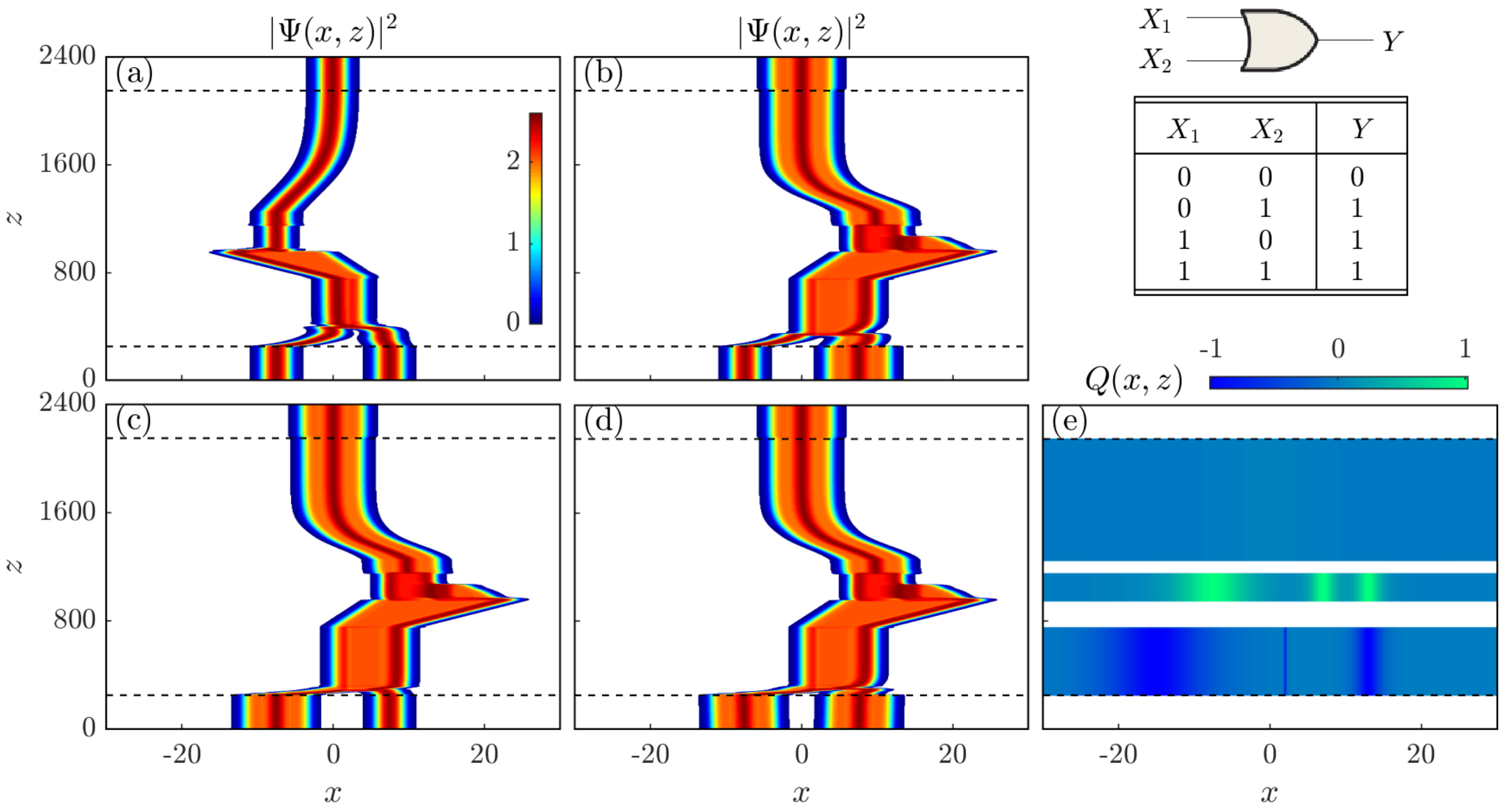}
\caption{OR gate. Truth table for a 2-input OR gate and its implementation on the dissipative soliton bits (a)-(d) due to externally applied potential $Q(x,z)$ with parameters \eqref{q_or} (e).}
\label{fig_5}
\end{figure*}

Below we demonstrate the implementation of an OR gate with two inputs in the framework of the model~\eqref{CQCGLE}-\eqref{BC} supporting the coexisting plain and composite pulses presented in Fig.~\ref{fig_1}. An OR gate produces a high level on the output when any of the inputs is high. The output is low only when all of the inputs are low. The operation of a 2-input OR gate is described in the truth table presented in Fig.~\ref{fig_5}, where the inputs are labeled $X_1$ and $X_2$, and the output is labeled $Y$. The implementation of an OR gate with two dissipative soliton inputs is demonstrated in Figs.~\ref{fig_5}(a)-(d), where each graph shows the intensity plot $|\Psi(x,z)|^2$ that represents the possible evolution of two input pulses into a single output under the control of potential~\eqref{Q} with the following parameters
\begin{align}
\label{q_or}
a_i\in\{250,950,1250\},~~~b_i\in\{750,1150,2150\},\nonumber\\
q_1(x)=-\mathrm{sech}\left(\frac{x+15}{4}\right)\nonumber\\ -\mathrm{sech}(x-13) - \mathrm{sech}(10x-20),\nonumber\\
q_2(x)=\mathrm{sech}\left(\frac{x+7.5}{3}\right)\nonumber\\ +\mathrm{sech}(x-7) +\mathrm{sech}(x-13),\nonumber\\
q_3=\frac{1}{20}\left[\mathrm{sech}\left(\frac{x}{5}\right)\right.\nonumber\\ \left.-\mathrm{sech}\left(\frac{x+19}{10}\right) -\mathrm{sech}\left(\frac{x-19}{10}\right)\right].
\end{align}

Again, we stress that the 2-input OR gate presented in Figs.~\ref{fig_5}(a)-(d) has been implemented due to the proper chosen parameters \eqref{q_or} of the potential~\eqref{Q}. Two-dimensional distribution of potential $Q(x,z)$ with parameters \eqref{q_or} is plotted in Fig.~\ref{fig_5}(e). Being applied this potential induces the plain pulse output only when both inputs are the plain pulses as shown in Fig.~\ref{fig_5}(a). On the other hand, when any of two inputs (including both of them) is the composite pulse, the output is the composite pulse [Figs.~\ref{fig_5}(b)-(d)]. Therefore, looking at the plain and composite pulses as Boolean variables whose values are respectively either binary $0$ or binary $1$ we conclude that Figs.~\ref{fig_5}(a)-(d) also implement the basic rules for Boolean addition.

The applied potential~\eqref{Q} with parameters~\eqref{q_or} performs the OR gate in three control stages, which are shown in Fig.~\ref{fig_5}(e). First of all, we apply $q_1(x)$ to transit four different pairs of input pulses to so-called \textit{moving} pulses \cite{Afanasjev_PRE_1996,PD_2019}. The asymmetrical profile of $q_1(x)$ is properly chosen to transit two input plain pulses to the moving pulse with a transverse drift along the negative direction of the $x$ axis [Fig.~\ref{fig_5}(a)], while all other input combinations are transited to the moving pulse with the opposite transverse drift [Figs.~\ref{fig_5}(b)-(d)]. The second stage succeeds the first one after some delay. During that delay the moving pulses are released of the potential influence that allows them to freely travel along the $x$ axis as seen in Figs.~\ref{fig_5}(a)-(d). At some moment the pulses traveling in opposite directions get sufficient spatial separation between them that allows us to apply potential $q_2(x)$ to transit the left shifted waveform to the plain pulse [Fig.~\ref{fig_5}(a)], while the right shifted waveforms are transited to the composite pulse [Figs.~\ref{fig_5}(b)-(d)]. Finally, having applied the weak potential $q_3(x)$ we symmetrically arrange the plain and composite pulses around the point $x=0$ and complete the implementation of the OR gate. 

\begin{figure*}[htbp]
\centering
\includegraphics[width=\linewidth]{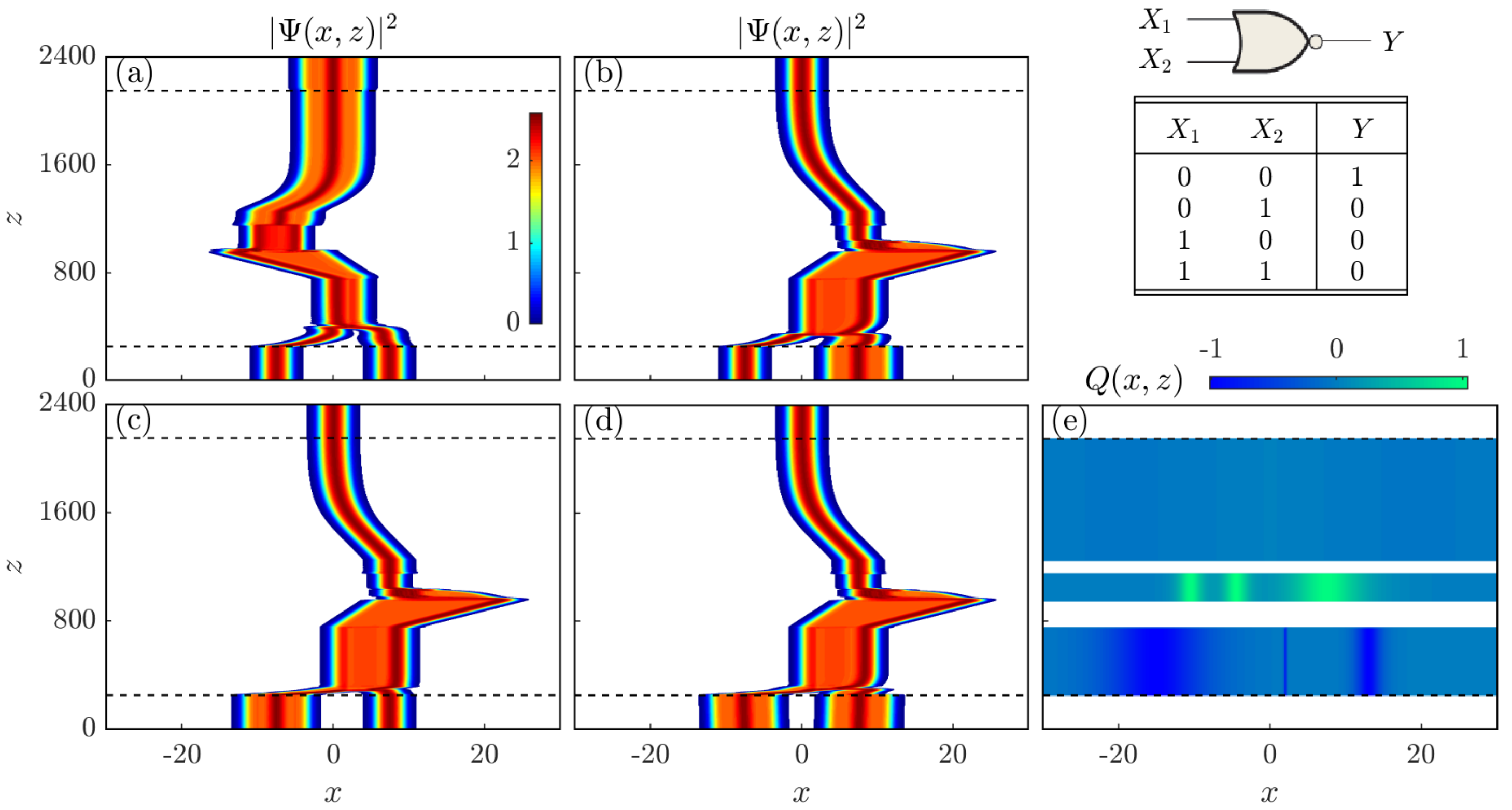}
\caption{NOR gate. Truth table for a 2-input NOR gate and its implementation on the dissipative soliton bits (a)-(d) due to externally applied potential $Q(x,z)$ with parameters \eqref{q_or}, where the second control manipulation is replaced by Eq.~\eqref{q_nor} (e).}
\label{fig_6}
\end{figure*}

It is logical to note that if during the second stage we transit the left (right) shifted waveform to the composite (plain) pulse we implement a 2-input NOR gate, which is the same as the OR except the output is inverted (see the truth table in Fig.~\ref{fig_6}). Therefore, we replace the potential parameter $q_2(x)$ in Eqs.~\eqref{q_or} by the following one
\begin{align}
\label{q_nor}
q_2(x)=\mathrm{sech}(x+10.5)+\mathrm{sech}(x+4.5)\nonumber\\ +\mathrm{sech}\left(\frac{x-7.5}{3}\right),
\end{align}
to get the potential suitable for the implementation of a 2-input NOR gate on dissipative soliton bits. The implementation of this NOR gate is shown in Fig.~\ref{fig_6}. All plots in Fig.~\ref{fig_6} are similar to those in Fig.~\ref{fig_5}. Indeed, the graphs in Figs.~\ref{fig_6}(a)-(d) show the two-dimensional intensity plots $|\Psi(x,z)|^2$, while Fig.~\ref{fig_6}(e) depicts the spatial distribution of control potential $Q(x,z)$ with parameters~\eqref{q_or}, where the parameter $q_2(x)$ is replaced by Eq.~\eqref{q_nor}. 

\subsection{\label{XOR}XOR and XNOR gates}
\begin{figure*}[htbp]
\centering
\includegraphics[width=\linewidth]{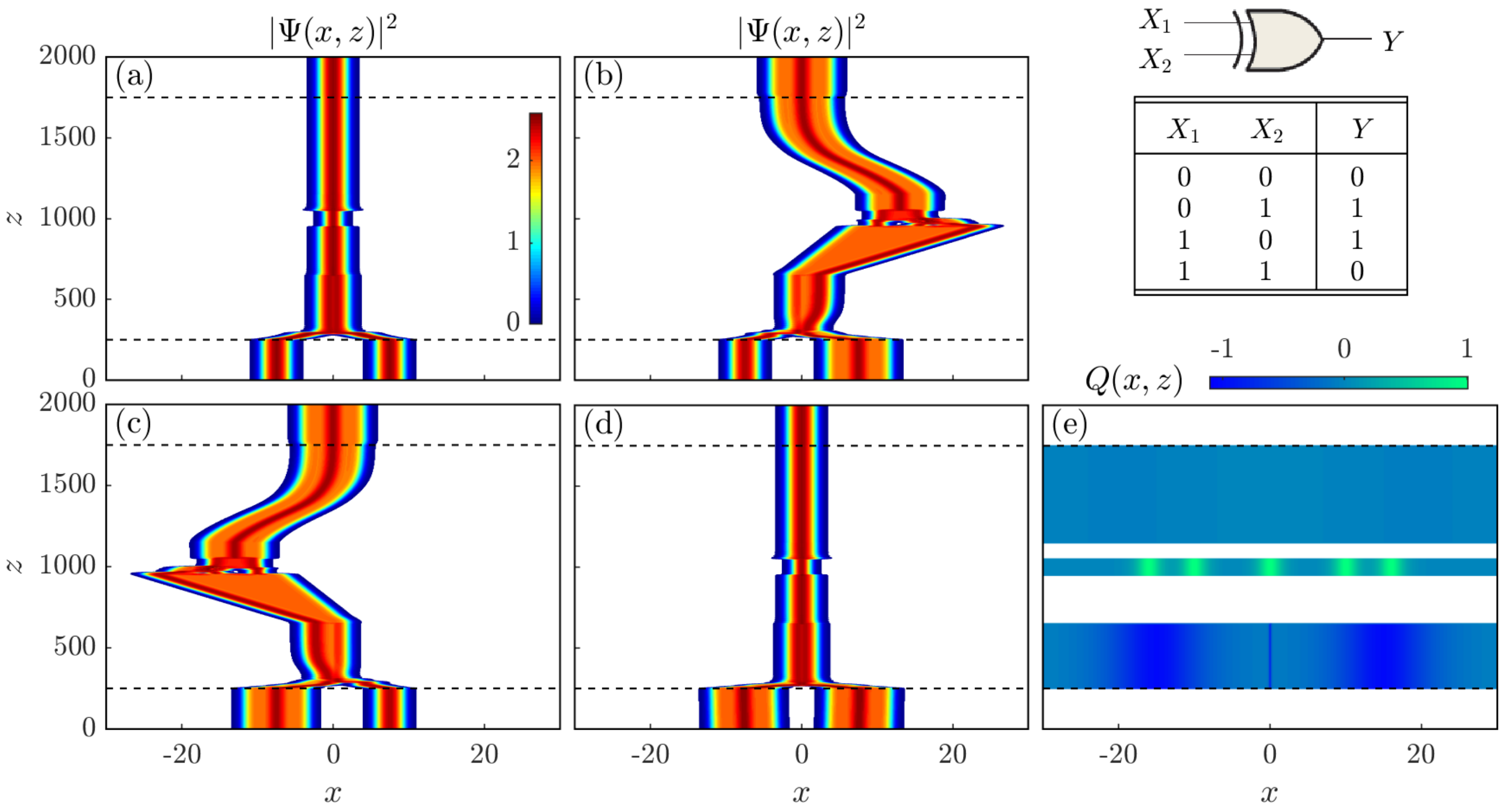}
\caption{XOR gate. Truth table for an exclusive-OR gate and its implementation on the dissipative soliton bits (a)-(d) due to externally applied potential $Q(x,z)$ with parameters \eqref{q_xor} (e).}
\label{fig_7}
\end{figure*}

Finally, we discuss the implementation of an XOR gate, which performs modulo-2 addition. Its operation is summarized in the truth table shown in Fig.~\ref{fig_7}. To implement an XOR gate on dissipative soliton bits we again employ the controllable model~\eqref{CQCGLE}-\eqref{BC} supporting the same plain and composite pulses (Fig.~\ref{fig_1}) as we used above. In Figs.~\ref{fig_7}(a)-(d) we demonstrate the four possible input combinations and the resulting outputs for the XOR gate implemented on the plain and composite pulses, where in each graph we plot the evolution of soliton intensities $|\Psi(x,z)|^2$. These simulations of the XOR gate have been performed using control potential~\eqref{Q} with the parameters specified as follows
\begin{align}
\label{q_xor}
a_i\in\{250,950,1150\},~~~b_i\in\{650,1050,1750\},\nonumber\\
q_1(x)=-\mathrm{sech}(10x)\nonumber\\ -\mathrm{sech}\left(\frac{x+15}{4}\right) -\mathrm{sech}\left(\frac{x-15}{4}\right),\nonumber\\
q_2(x)=\mathrm{sech}(x+16) +\mathrm{sech}(x+10) \nonumber\\ +\mathrm{sech}(x) +\mathrm{sech}(x-10) +\mathrm{sech}(x-16),\nonumber\\
q_3=\frac{1}{20}\left[\mathrm{sech}\left(\frac{x}{5}\right)\right.\nonumber\\ \left.-\mathrm{sech}\left(\frac{x+19}{10}\right) -\mathrm{sech}\left(\frac{x-19}{10}\right)\right].
\end{align}
The two-dimensional spatial distribution of this control potential is illustrated in Fig.~\ref{fig_7}(e).

\begin{figure*}[htbp]
\centering
\includegraphics[width=\linewidth]{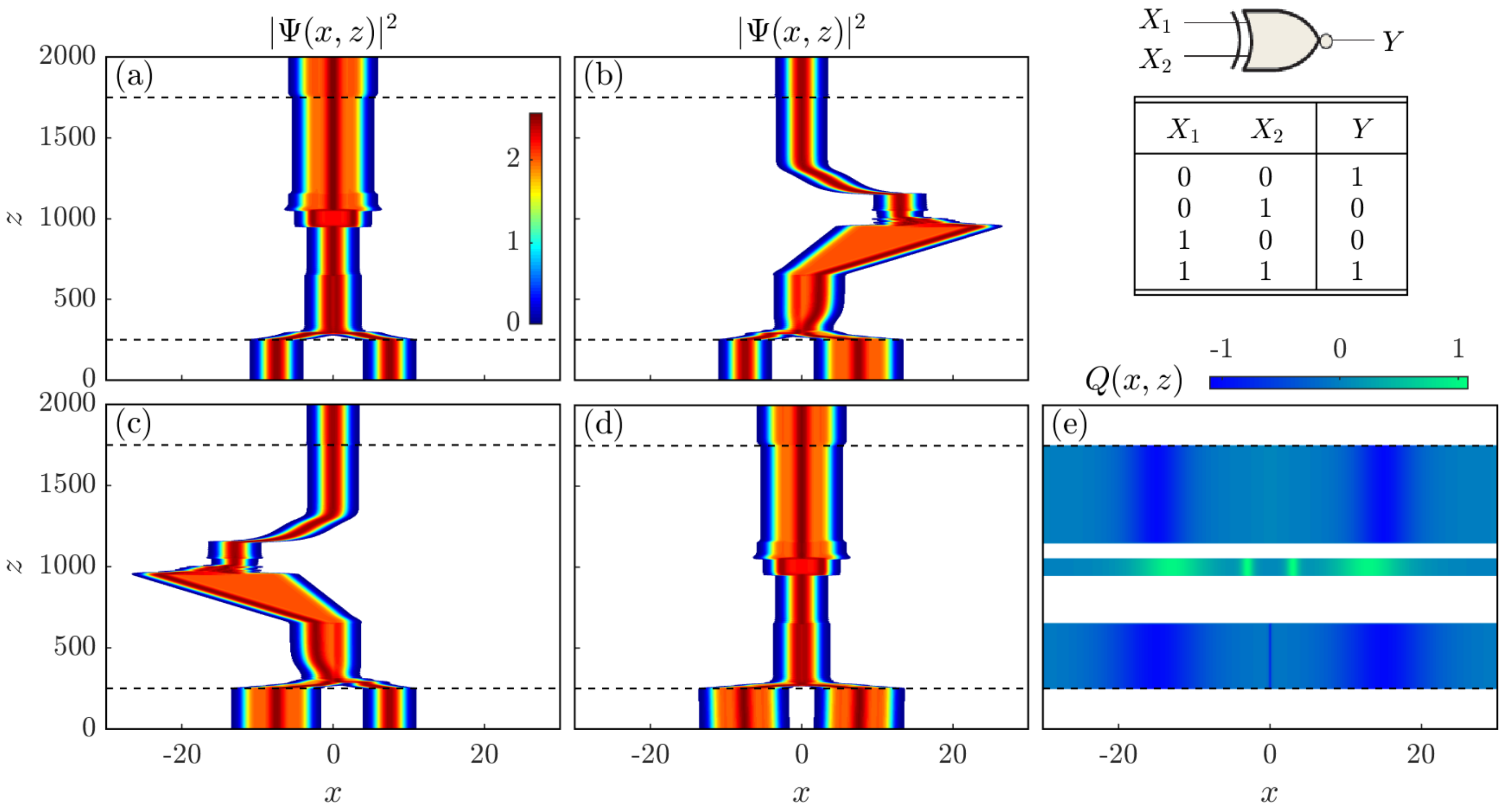}
\caption{XNOR gate. Truth table for an exclusive-NOR gate and its implementation on the dissipative soliton bits (a)-(d) due to externally applied potential $Q(x,z)$ with parameters \eqref{q_xor}, where the last two control manipulations are replaced by Eqs.~\eqref{q_xnor} (e).}
\label{fig_8}
\end{figure*}

Since the high output level occurs in an XOR gate only when the inputs are at opposite levels we firstly apply the potential $q_1(x)$ that selectively transits the pairs of input pulses depending on whether the same or different pulses are launched to the gate. Indeed, two plain pulses [Fig.~\ref{fig_7}(a)] as well as two composite pulses [Fig.~\ref{fig_7}(d)] are transited by the potential~\eqref{Q} with parameters \eqref{q_xor} to the plain pulse, while the combinations of plain - composite [Fig.~\ref{fig_7}(b)] and composite - plain [Fig.~\ref{fig_7}(c)] pulses are transited to the moving pulses with positive and negative drifts along the $x$ axis, respectively. Between first and second control manipulations is some lag leading to the significant displacements of moving pulses along the $x$ axis as illustrated in Figs.~\ref{fig_7}(b) and (c). After that, we apply the potential $q_2(x)$ to transit the shifted moving pulses to the composite pulse [Figs.~\ref{fig_7}(b) and (c)] and keep the plain pulses unchanged [Figs.~\ref{fig_7}(a) and (d)]. Finally, we apply the weak potential $q_3(x)$ to arrange all the pulses around the point $x=0$.

Thus, we has completed the implementation of the XOR gate and found a simple way to implement the XNOR gate whose outputs are opposite to those of the XOR gate, as summarized in the truth table for an XNOR gate shown in Fig.~\ref{fig_8}. In fact, to implement the XNOR gate we again use potential~\eqref{Q} with parameters \eqref{q_xor} except for the last two control manipulations, which are now chosen in the following form
\begin{align}
\label{q_xnor}
q_2(x)=\mathrm{sech}(2x+6) +\mathrm{sech}(2x-6) \nonumber\\ +\mathrm{sech}\left(\frac{x+13}{3}\right) +\mathrm{sech}\left(\frac{x-13}{3}\right),\nonumber\\
q_3(x)=\frac{1}{20}\mathrm{sech}(x) \nonumber\\
-\mathrm{sech}\left(\frac{x+15}{2.5}\right) -\mathrm{sech}\left(\frac{x-15}{2.5}\right).
\end{align}

Having performed the first stage and being waited for some lag between first and second manipulations we proceed to the second stage. During the second stage we now apply potential $q_2(x)$ from Eqs.~\eqref{q_xnor} to transit the plain pulses to the composite pulse [Figs.~\ref{fig_8}(a) and (d)] and the laterally shifted (moving) pulses to the plain pulse [Figs.~\ref{fig_8}(b) and (c)]. Finally, we apply potential $q_3(x)$ from Eqs.~\eqref{q_xnor} to center the shifted pulses with respect to the point $x=0$ and complete the implementation of XNOR gate. We replace the last control manipulation $q_3(x)$ to move the shifted plain pulses [Figs.~\ref{fig_7}(b) and (c)] faster than we moved the corresponding composite pulses in Fig.~\ref{fig_7}.

\section{\label{Concl}Conclusions}
In our numerical simulations presented in Figs.~\ref{fig_2}-\ref{fig_8} we have demonstrated the implementation of basic logic gates, where two logic levels are represented by two stationary dissipative solitons with distinguished waveforms and spectra. The simulations have been carried out in the framework of the one-dimensional cubic-quintic CGLE with a potential term~\eqref{CQCGLE}. On the one hand this equation accounts for the most important features of dissipative solitons while admitting existence of a wide range of sophisticated solutions on the other hand. That makes it one of the basic mathematical models of dissipative solitons in many applications. Particularly, Eq.~\eqref{CQCGLE} admits coexistence of two stationary solutions in the from of plain and composite pulses (Fig.~\ref{fig_1}), which are used here to represent the low and high logic levels, respectively. Moreover, Eq.~\eqref{CQCGLE} contains the external potential $Q(x,z)$, which we apply to control evolution of solitons within the system. In fact, the potential has a vital impact on the soliton dynamics and plays the most important role in the implementation of logic gates. In each simulation we applied potential~\eqref{Q} with some parameters properly chosen to implement a given logic gate. These particular parameters are listed in Eqs.~\eqref{q_not}-\eqref{q_xnor}, where the transverse profiles $q_i(x)$ have been chosen in the form of combinations of scaled $\mathrm{sech}(x)$ functions. However, they contain all the significant properties of the potentials suitable to implement the logic gates. In general, an appropriate potential implements a logic gate in three basic control stages. This operation can be stated as follows:
\begin{itemize}
\item The potential selectively transforms and shifts along the transverse direction the pairs of input pulses depending on their input combinations. In other words, during the first stage we perform the transverse spatial selection of input solitons, i.e. different combinations of input pulses get different lateral shifts.
\item The potential transits each spatially separated pulse to the proper output. During the second stage some of the pulses can also be shifted along the transverse direction.
\item The potential gradually performs lateral shift of the output pulses to arrange all the outputs around the same point. 
\end{itemize}
We have to note that each of those stages can consist of one or several particular control manipulations. For example, the first stage for all the gates discussed here (Figs.~\ref{fig_2}-\ref{fig_8}) consists of two control manipulations. For the AND (Fig.~\ref{fig_2}) and NAND (Fig.~\ref{fig_3}) gates the first stage has been performed by two functions $q_1(x)$ and $q_2(x)$ from Eqs.~\eqref{q_and}, but the same stage of all other gates (Figs.~\ref{fig_4}-\ref{fig_8}) has been performed by the corresponding functions $q_1(x)$ and subsequent control manipulations with zero transverse functions as shown in Figs.~\ref{fig_4}(e)-\ref{fig_8}(e). On the other hand, the third stage for each of the logic gates has been performed by a single control manipulation. For the AND and NAND gates it is function $q_5(x)$ in Eqs.~\eqref{q_and} and Eqs.~\eqref{q_nand}, while for the other gates it is function $q_3(x)$ in Eqs.~\eqref{q_or}-\eqref{q_xnor}, respectively. 

In this paper we have demonstrated a numerical implementation of the basic logic gates with two inputs applying external potentials to get control over soliton waveforms. Particularly, we have found the potentials, which allow us to get the proper output soliton waveforms depending on a given input. Due to our numerical simulations we have shown that such potentials should contain three vital control manipulations over input pulses: transverse spatial selection, waveform transition, and output arrangement as summarized above. 

Traditionally, digital electronics is built on universal NAND gates, whose combinations can be used to produce any logic function. However, in this case, we have a planar technology, where switching between pulses across tracks is difficult. Therefore, we have considered various logic gates implemented in the planar form.

This approach can also be used to implement other logic gates with two or more inputs as well as to model such optical devices as splitters, demultiplexers, cellular automata, etc., where the dissipative solitons are employed as logic levels (bits). Moreover, the ideas discussed here can be useful for experimental studies on dissipative optical solitons.

\section*{\label{ack}Acknowledgments}
The authors are grateful for support from Jilin University, China.

\bibliographystyle{model1-num-names}
\bibliography{Soliton}

\end{document}